# Large Area Fabrication of Semiconducting Phosphorene by Langmuir-Blodgett Assembly

Harneet Kaur[1], Sandeep Yadav[2], Avanish. K. Srivastava[1] Nidhi Singh[1], Jörg J. Schneider[2], Om. P. Sinha[3], Ved V. Agrawal[1,+], and Ritu Srivastava[1,+,*]

[1]National Physical Laboratory, Council of Scientific and Industrial Research, Dr. K. S. Krishnan Road, New Delhi 110012, India

[2]Technische Universität Darmstadt, Eduard-Zintl-Institut für Anorganische und Physikalische Chemie L2 I 05 117, Alarich-Weiss-Str 12, 64287 Darmstadt, Germany

[3]Amity Institute of Nanotechnology, Amity University, Sector 125, Noida, Uttar Pradesh 201313, India

**Abstract**

Phosphorene is a recently new member of the family of two dimensional (2D) inorganic materials. Besides its synthesis, it is of utmost importance to deposit this material as thin film in a way that represents a general applicability for 2D materials. Although a considerable number of solvent based methodologies have been developed for exfoliating black phosphorus, so far there are no reports on controlled organization of these exfoliated nanosheets on substrates. Here, for the first time to the best of our knowledge, a mixture of N-methyl-2-pyrrolidone and deoxygenated water is employed as a subphase in Langmuir-Blodgett trough for assembling the nanosheets followed by their deposition on substrates and studied its field-effect transistor characteristics. Electron microscopy reveals the presence of densely aligned, crystalline, ultra-thin sheets of pristine phosphorene having lateral dimensions larger than hundred of microns. Furthermore, these assembled nanosheets retain their electronic properties and show a high current modulation of $10^4$ at room temperature in field-effect transistor devices. The proposed technique provides semiconducting phosphorene thin films that are amenable for large area applications.

**Introduction:**

Two dimensional (2D) inorganic materials form a class of nano-materials, having a thickness of a few atomic layers. The success of graphene[1-3], the most widely known 2D material in research community, is an inspiration in discovering this new class in material science. The credit for its success is its strength[4], extremely light weight[5], highly flexibility with high conductivity towards heat and electricity[6-9]. However, from the perspective of semiconductor industry, graphene has a drawback, absence of pristine bandgap[2], resulting in poor current modulation ratio in field-effect transistors (FET)[2,10]. Apart from the ongoing research in introducing the bandgap in graphene[10-13], a new momentum has emerged when a group of researchers shown the presence of a direct pristine bandgap[14,15] (1.8 eV) in single layer molybdenum disulphide ($MoS_2$). Consequently, when this single layer is used in FET, it overcomes the existing problem[16] with graphene by offering a high current modulation[17] ratio of $10^8$. This has resulted in the creation of immense interest in physicists and chemists to explore the family of transition metal dichalcogenides (TMDs) to which $MoS_2$ belongs and as a result, a new family of 2D material (graphene, $MoS_2$, GaS, GeS, $TaS_2$, $WS_2$, $HfS_2$, CdS, $SnS_2$ etc) is born[18].

Recently, a new member has been introduced to the family of 2D materials, black phosphorus (BP); the most stable allotrope of phosphorus[19-21]. The presence of its stacked layered structure and weak van der Waals interlayer forces makes it possible to cleave it along the individual layers to produce atomic thin layers called phosphorene. The successful fabrication of phosphorene based FET[22,23] in 2014 has resulted in a rapid progress towards unrevealing its novel properties including its anisotropy in conductivity[24-26], young modulus[27,28], phonon-electron interaction[29,30] apart from its promising direct bandgap which can be tuned from 0.3 eV in bulk to 1.5 eV in single layer[19-23]. Even few layers of phosphorene hold promise for FETs as they provide moderate current modulation ratios with high charge carrier injection[22,23,31-33]. Currently, fabrication of phosphorene consists of several top-down methods whereas the bottom-up approaches like chemical vapor deposition (CVD) or hydrothermal method are still blank. This may be due to the high reactivity of phosphorene[34,35]. Very recently, substrates effects have been predicted theoretically where a mediocre interaction of phosphorene with substrate can result in its stable epitaxial monolayer growth[36] but no experiments were demonstrated yet. Among the top-down methods, the mostly widely used is mechanical exfoliation[22,23,32] but it is limited to only laboratory use. Another method is plasma assisted thinning[37] of BP by thermal ablation requiring laser raster scanning making it challenging for scale up applications. Also, modified mechanical exfoliation employing $Ar^+$ plasma thinning is also demonstrated but is only suitable for small scale applications[38]. Recently, Li *et.al* [39] demonstrated a method of growing red allotrope of phosphorus on flexible substrate followed by its conversion into black allotrope. However, this method has resulted in large area fabrication but the minimum thickness achieved is 40 nm. Therefore, for bulk production of thin layers of phosphorene and other 2D materials, liquid phase exfoliation methods are very promising[40-44]. However, the yield is high but it produces exfoliated atomic thin layers suspended in solvent and not on the solid support. So if these two-dimensional exfoliated nanosheets suspended in solvent can be ordered, and assembled systematically on the substrate, they will offer a versatile, inexpensive, mass production method for the large area fabrication of thin films of layered materials.

A lot of research has been devoted in past few years into the area of organizing graphene and graphene oxide suspended in solvent into extended spatial arrangements[45-47] but no established protocol other than drop-cast[42,43,48] has been reported till date into the area of organizing phosphorene nanosheets on substrates. In this article, we propose a novel environment benign Langmuir-Blodgett (LB) method to assemble exfoliated phosphorene nanosheets with dense packing on the substrate. A typical LB process consists of water subphase onto which amphiphilic molecules suspended in volatile solvent are injected. The solvent is evaporated resulting in trapping of molecules on air/water interface resulting in the formation of LB monolayer. However, in case of phosphorene, surface easily gets oxidized in presence of water as well ambient oxygen[34,35,37]. Therefore, we modify this typical method by using a mixture of N-methyl-2-pyrrolidone (NMP) with deoxygenated water as subphase medium. This has resulted in the assembly followed by deposition of un-oxidized pristine phosphorene thin films which retain their semiconducting properties as demonstrated by fabrication of FET. Thus, our proposed assembly procedure not only holds promise to provide large area semiconducting phosphorene thin films but also provides a possibility to deposit other 2D materials which are sensitive to oxygen.

**Results and Discussion:**

The red allotrope of phosphorus was first converted into its black allotrope as discussed in the method section. The structure, morphology and elemental analysis of the as-prepared crystal were quantified through X-ray diffraction (XRD), Raman spectroscopy, scanning electron microscopy (SEM) and energy-dispersive X-ray spectroscopy (EDX). Figure 1a shows X-ray diffraction pattern of the crystal which reveals the presence of high intensity peaks corresponding to miller indices (*hkl*) 020, 040 and

060 planes as assigned by the JCPDS number 76-1961. The *d* spacing between the *hkl* planes has been calculated that matches well with the reported values of bulk BP[38]. Further, the Raman vibrational modes corresponding to $A^1_g$, $B_{2g}$ and $A^2_g$ at 362.3 cm$^{-1}$, 438.3 cm$^{-1}$ and 466.2 cm$^{-1}$ respectively (Figure 1b) confirms the structure of BP[37]. In Figure 1c, the SEM image of the BP crystal clearly shows its layered structure and presence of phosphorus atoms was confirmed by the EDX (inset of Figure 1c).

The delamination of the van der Waals bonded BP crystal into single and few layers of various sizes were accomplished by solvent assisted ultrasonication method in a sealed bottle followed by centrifugation at various speed (r.p.m). We used NMP as a solvent. The Tauc plot (Figure 1d) was calculated using the optical absorbance spectra (supplementary information, Figure S1) of the as-prepared suspension. The linear relationship between *(αhv)$^2$* and photon energy *(hv)* confirms that the bandgap is direct. By fitting the linear part with a straight line, the high energy transition was coming out to be 2.2 eV. Since, bulk BP has a higher energy transition at 1.9 eV thus, increase in higher energy transition is suggesting suspension of exfoliated phosphorene which is in good agreement with the previous experimental work[50]. Further, atomic force microscopy (AFM) on the drop-cast BP films in Figure 1e revealed the thickness of these nanosheets varies between 2-4 nm as predicted by its height profile (inset of Figure 1e). Previous AFM studies showed the thickness of single layer phosphorene on silicon dioxide grown silicon substrates (SiO$_2$/Si) substrates is 0.9 nm[23] for mechanical exfoliated flakes, which is more than its theoretical thickness. Also, the height of single layer graphene deposited by LB assembly as predicted by AFM is around 1 nm[47] which is more than its theoretical value[2] 0.35 nm. Therefore, AFM results confirm that the nanosheets predominantly consist of two to five layer phosphorene. Also, the presence of sharp edges as well as absence of bubble formation[35,37] on the surface of nanosheets confirms that the nanosheets were not oxidized. Thus, the pristine phase was preserved during its exfoliation which is mandatory for its electronic applications. However, we found that NMP is a good solvent for suspending the phosphorene sheets only if it was stored in a sealed bottle and kept in mild vacuum (~$10^{-2}$ mbar). In ambient conditions, the color of the suspension gradually fades over time which was attributed to the oxidation of the exfoliated nanosheets (supplementary information, Figure S2). Nevertheless, it is highly desirable that the suspension is stable and retains its un-oxidized phase for Langmuir assembly.

The phosphorene suspension in NMP was spread drop wise (typically 0.5 ml) on the subphase of LB trough. The subphase used for the assembly was a mixture of NMP and deoxygenated deionized water in the ratio of 1:10. It has been reported that there is a strong hetero-association of NMP with water molecules[51] due to the formation of polymeric species of the type (N,N-disubstituted amide·$3H_2O$)$_n$ or (N-substituted amide·$2H_2O$)$_n$.[34] This results in the more compactness of NMP, thereby, the subphase than in the pure state. Further, it was observed that NMP acts as an encapsulation layer preventing ambient degradation of phosphorene sheets during LB assembly. Thus the presence of NMP in the subphase is a crucial parameter for assembling the un-oxidized compact arrangement of phosphorene nanosheets as confirmed by the Raman spectroscopy[35] (supplementary information, Figure S3). The barriers were compressed slowly while the surface pressure was monitored with a Wilhelmy plate. The area-pressure isotherm (supplementary information, Figure S4) confirms the floating nature of phosphorene nanosheets. Due to the presence of NMP molecules in the suspension, the compression starts with a non-zero surface pressure. The phosphorene nanosheets were transferred using the vertical lift-off procedure onto $SiO_2$/Si substrates and carbon coated copper grids at a surface pressure of 40 mN/m (supplementary information, Figure S5).

Field emission scanning electron microscope (FESEM) was used to characterize the morphology of LB assembled nanosheets on $SiO_2$/Si substrates. Figure 2a shows a representative, low-magnification FESEM image (0.3 X 0.3 $mm^2$) revealing a high density deposition of small nanosheets (S-Ex BP, centrifuged at 10,000 r.p.m). Such a compact spatial arrangement of aligned nanosheets over such a large area cannot be achieved through drop-cast method as illustrated above. It can only be used where the small microscopic area (in microns) is of interest, hence cannot be used to produce thin films of phosphorene. The magnified image (Figure 2b), indicates nanosheets were flat and have lateral dimensions in the order of few microns. Figure 2c shows its AFM image, reveals thickness of these nanosheets varies from 3-5 nm. No flake having thickness more than 5 nm was found is an indicative of the presence of few layer phosphorene (supplementary information, Figure S6). Also this range of height was comparable with the thickness of the nanosheets deposited by drop-cast method ensures that during the assembly procedure, the nanosheets were neither oxidized nor underwent any degradation. Further, LB films made by using large nanosheets suspension (L-Ex BP, centrifuged at 3000 r.p.m) shows an enrichment of ultra-large nanosheets (>10,000 $\mu m^2$, supplementary information, Figure S7) on the substrate as depicted in Figure 2d which represents a low magnification image (0.8 X 0.8 $mm^2$). The continuity in the nanosheets is clearly shown in its magnified image (Figure 2e) over several microns. However, small dark contrast features (marked by circle in Figure 2d) were attributed to be formed from the curling or rolling of these large nanosheets during LB assembly under high surface pressure. Figure 2f shows the AFM image of L-Ex BP, which shows that these large nanosheets have thickness ~ 4 nm. Thus, LB assembly technique can result in the production of controllable deposition of large area phosphorene nanosheets with high reproducibility.

Further, high resolution transmission electron microscopy (HRTEM) was employed to characterize the LB assembled nanosheets on TEM grids. A detailed electron microscopy study has led to the revelation of several micro structural features in real and reciprocal space. In general a uniform microstructure of phosphorene was discerned throughout in the studied specimen (Figure 3a). Inset (i) in Figure 3a elucidates mingling of ultra-fine sheets of phosphorene while inset (ii) shows an enlarged view of few mingled sheets as shown in inset (i), exhibiting the evolution of atomic planes with a regular spacing of about 0.51 nm and 0.18 nm with the miller indices (*hkl*) of 020 and 112, respectively (crystal structure: orthorhombic, lattice constants: a = 0.331 nm, b = 1.029 nm, c = 0.4302 nm, space group: Cmca, reference: JCPDS card no. 76-1961). A corresponding selected area electron diffraction pattern (SAEDP) from the aggregate of phosphorene nanosheets (inset (i) in Figure 3a) shows the presence of a set of Debye rings in reciprocal space, as displayed in Figure 3b. The Debye rings in Figure 3b corresponds to important planes of orthorhombic crystal structure of phosphorene with *hkl* indices as 020, 040, 111 (marked on Figure 3b). A set of SAEDPs recorded from the phosphorene nanosheet (Figure 3a) elucidates single crystalline electron diffraction patterns, as depicted along [012], [210] and [110] zone axes of orthorhombic crystal structure of phosphorene in Figure 3c,d,e. The SAEDPs (Figure 3c,d,e) clearly reveals that the individual sheets of phosphorene were well crystallized with the organization of atomic planes in the single crystalline nature with no evidence of structural disorder or oxidation. Figure 3f displays an illustrative example of atomic scale image of phosphorene with the stacking of planes (hkl: 040) corresponding to inter-atomic separation of 0.26 nm. In Figure 3g, a boundary between two phosphorene sheets (I and II) has been marked with a set of arrows, although both the sheets are aligned with an inter-atomic separation of 0.26 nm. It is interesting to note that a magnified view of phosphorene sheet (marked with white dotted region in Figure 3g) discerns a honey-comb arrangement of phosphorene (inset (iii) in Figure 3g) in the microstructure.

To elucidate the semiconducting properties of the material, a FET device was fabricated on doped silicon substrates, having 230 nm of silicon dioxide as a dielectric with gold patterning on top (Figure 4a). The Si-substrate acts as a gate electrode and controls the channel current between the gold electrodes on top which acts as source and drain. The L-Ex

BP nanosheets were assembled on these substrates resulting in a conducting channel between 5 μm wide gold electrodes (Figure 4b). Further, high magnification optical image of the channel shows bridging between source and drain through a phosphorene sheet (Figure 4c).The thickness of the deposited phosphorene was determined to be ~ 4 nm as revealed by AFM (Figure 4d). The switching behavior of the FET at room temperature is presented in Figure 4e. The back gate voltage $V_{GS}$ was swept from -10V to 0V, keeping drain voltage $V_{DS}$ at -1V. The maximum on state drain current $I_{DS}$ is $10^2$ μA/μm while off-state current is $10^{-2}$ μA/μm enabling a high on-off current modulation ratio of $10^4$. Figure 4f represents the $I_{DS}$ versus $V_{DS}$ graph for different step voltages of $V_{GS}$ which shows that current in few layer phosphorene FET can be controlled by providing a suitable gate voltage. Comparing this $I_{on}/I_{off}$ ratio with the previous reported values[23, 47, 52], it is evidenced that LB assembly is a suitable and convenient method for the fabrication of superior quality phosphorene thin films. In conclusion, synthesis of black allotrope of phosphorus followed by its effective exfoliation was achieved. Langmuir Blodgett technique using NMP mixed with deoxygenated water as assembling medium, results in the systematic deposition of un-oxidized phosphorene nanosheets. The direct assembly of ultra-large nanosheets onto gold patterned $SiO_2$/Si substrates allows a quick, straightforward fabrication of FET devices. The resulting FET device offers a high current on-off ratio of $10^4$ at room temperature. Thus, the ability and simplicity to fabricate large area, electronic quality phosphorene using this above mentioned wet route could be further extended as a general strategy to assemble other graphene analogues materials which are sensitive to ambient oxygen. Further, this method will provide a cost-effective solution for fabrication of large area phosphorene films with performance superior to conventional methods.

**Methods**

Crystal Growth

Bulk black phosphorous crystals was synthesized by sublimation of red phosphorous (Sigma Aldrich, 99.999 \%) in the presence of tin and tin iodide (IV) (ABCR GmbH 99.999 \%) in an evacuated (approx. $3\times10^{-3}$ mbar) quartz ampule having inner diameter 1 cm and 0.25 mm wall thickness. The tube containing the materials was kept in oven and the temperature has been raised slowly upto $695^{o}C$ at the rate of $7^{o}C$ per minute and kept constant for 6 hours followed by slow cooling to $550^{o}C$ in 8 hours. The material remained in the oven for further 72 hours at $550^{o}C$, followed by cooling the as-prepared crystals to room temperature for characterization. X-ray diffraction (XRD), Raman spectroscopy, scanning electron microscopy (SEM) and energy dispersive analysis of X-Rays (EDX) was performed using Rigaku miniFlex 600, Horiba scientific (LabRAM HR Raman, Green Laser-514.5 nm), Philips XL-30 FEG and XL30-939-2.5 CDU-LEAP-detector respectively.

Exfoliation

As-prepared bulk BP crystals (10 mg) were dispersed in anhydrous NMP, (10ml, Sigma-Aldrich) and kept in a sealed vial of volume 20 ml. Further, these sealed vials were wrapped with parafilm before placing it into elmasonic TI-H bath sonicator. The dispersion was sonicated in DI water for 4 hours at 25 KHz. Temperature of DI water in ultrasonic bath was kept at $30^{o}C$. The color of the suspension gradually changes from black to brownish yellow after four hours, yielding a suspension of few-layers phosphorene nanosheets in NMP. The suspension was centrifuged at 3000 r.p.m for 30 min in a spinwin micro centrifuge (Tarsons) to remove the un-exfoliated material and the supernatant was decanted, and subjected to UV-Visible absorption measurement using Shimadzu, UV-2401. Atomic force microscopy studies (SOLVER P47 PRO-NT-MDT) were carried out on drop cast suspension of phosphorene on $SiO_2$ grown Si substrates. The suspension was further centrifuged at 10,000

r.p.m for 30 min, to size separate the smaller and larger nanosheets of BP. The supernatant containing the smaller nanosheets named, S-Ex BP and the sediment containing the large nanosheets was dispersed in fresh NMP and named L-Ex BP. Both S-Ex BP and L-Ex BP were stored in sealed flasks under evacuated conditions (~$10^{-2}$ mbar).

Langmuir Blodgett Assembly

KSV-NIMA (Model: KN-1006, Langmuir trough) was used for the assembly of nanosheets of phosphorene. Prior to deposition, all the parts of LB trough were thoroughly cleaned by propanol, acetone and chloroform. The subphase used for the assembly was a mixture of anhydrous NMP and deoxygenated DI water in the ratio of 1:10. Initially, DI water (700 ml) in a 1 l flask was degassed by purging argon gas for half an hour to remove any dissolved oxygen gas. Thereafter, 70 ml of NMP was added and it was further purged with argon gas and stirred 30 min. The prepared degassed mixture of NMP and DI water was poured onto the LB trough and temperature of trough was maintained at 35°C. We have found that the mixture of NMP and DI water does not yield any surface pressure even after compressing the barriers and thus was used as subphase in all experiments. The suspensions of exfoliated phosphorene (S-Ex BP and L-Ex BP) were slowly injected by a glass syringe (~0.5 ml). The barrier compression was immediately begun afterwards. The barriers were compressed at the rate of 10 mm/min. The LB monolayer was transferred to TEM grids covered with amorphous carbon thin films (400 mesh), silicon substrates with 230 nm of silicon dioxide and n-doped (~3X $10^{17}$ $cm^{-3}$) silicon substrates with 230 nm silicon dioxide as a gate dielectric with gold electrodes on top (OFET chip, Fraunhofer), by vertically lifting at a speed of 1mm/min at a surface pressure of 40 mN/m. The wetting of substrate is a necessary condition for the compact deposition of phosphorene nanosheets by LB assembly hence, $SiO_2$/Si substrates being hydrophilic, was used in the experiment. It was further ensured that the experiments were completed in 20 min after injection of phosphorene suspension to avoid the ambient oxidation of nanosheets. All the grids and substrate were immediately transferred to vacuum oven ($10^{-5}$ mbar) and were maintained at $60^\circ$C for 6 hours for drying.

Characterization

Carl Zeiss, Supra 40 VP FE-SEM was used to characterize the morphology of LB films on $SiO_2$/Si substrates at an accelerating voltage of 3 KV. The AFM images were scanned in tapping mode at a frequency of 1 Hz using Pro P47 SOLVER, NT-MDT. The optical images were acquired using WITEC alpha300 R confocal microscope. A Tecnai G2 F30 STWIN, TEM was used for low and high resolution imaging of LB prepared TEM grids at 300 KV accelerating voltage. The electrical measurements were done using Keithley 4200 equipped with semiconductor characterization system (Summit 11000M, Probe station).

FET Characterization:

Due to contribution from several nanoflakes in electrical response, we are not able to comments on either current density or mobility of the device at this stage.

ACKNOWLEDGMENT

Growth of crystal and its characterization was sponsored by the LOEWE project STT funded by the state of Hesse at Technische Universität Darmstadt. Fabrication of devices and its characterization conducted is sponsored by UGC-SRF, UGC-DAE CSR IC/CRS-77 Indore and CSIR-TAPSUN NWP-55 project.

**Author Contributions**

H.K. designed the experiments, exfoliates the crystal, performed Langmuir-Blodgett assembly, UV-Visible absorption and Raman spectroscopy, AFM characterization, fabricates the devices and wrote the manuscript. S.Y. and J.J.S. grown the crystals and characterized it. A.K.S. performed TEM characterization and its analysis and N.S. performed FESEM characterization. J.J.S., O.P.S., V.V.A. and R.S. analyzed the results and edited the manuscript. All authors reviewed the manuscript. V.V.A. and R.S. contributed equally to this work.

Additional information

Supplementary information accompanies this paper; Competing financial interests: The authors declare no competing financial interests.

FIGURES

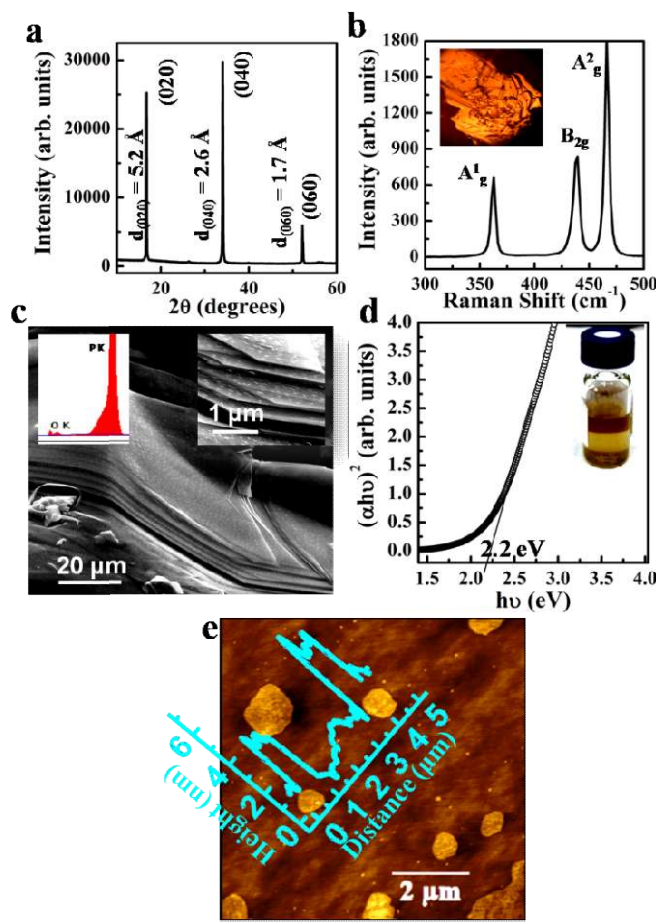

**Figure 1.** Characterization of BP crystal and its exfoliated suspension. (a) X-ray diffraction spectra of BP crystal. (b) Raman spectra of the BP crystal. Inset: Optical image of crystal. (c) Scanning electron microscopy of BP revealing its layered structure. Inset: EDX spectra of BP crystal (left) and magnified SEM revealing the presence of sharp edges (right). (d) Tauc plot of exfoliated BP suspension. Inset: Digital image of exfoliated BP suspension. (e) Atomic force microscopic image of the drop-cast nanosheets of exfoliated BP. Inset: height profile of phosphorene nanosheets.

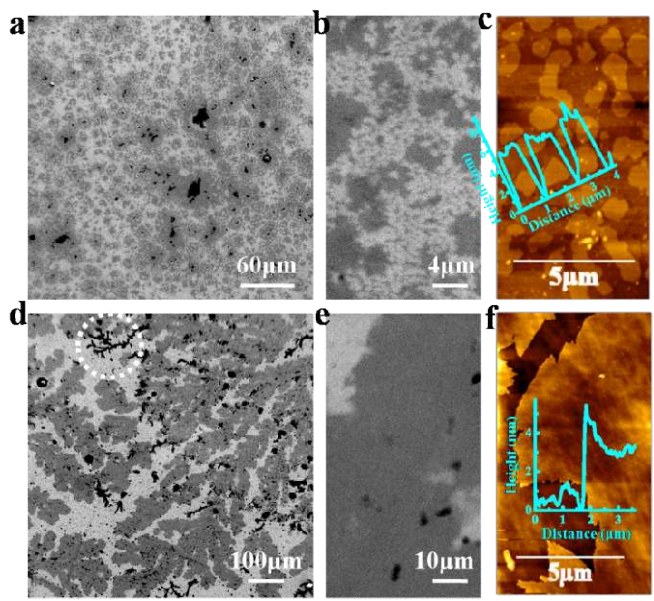

**Figure 2.** Surface morphology of LB assembled phosphorene nanosheets on SiO$_2$/Si substrate. (a) FESEM of small nanosheets (S-Ex BP) deposited at a surface pressure of 40 mN/m. (b) Magnified FESEM image S-Ex BP. (c) AFM of S-Ex BP. Inset: height profile of nanosheets. (d) FESEM of L-Ex BP nanosheets deposited at a surface pressure of 40 mN/m. (e) Magnified FESEM image of L-Ex BP. (f) AFM of L-Ex BP. Inset: height profile of nanosheets.

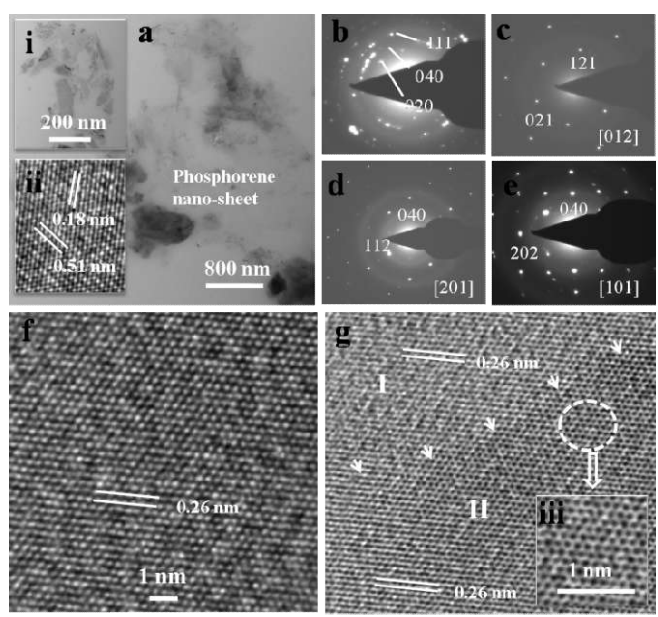

**Figure 3.** HRTEM of LB assembled phosphorene on TEM grids. (a) Thin sheets of phosphorene. Inset (i): Aggregate of thin sheets of phosphorene, (ii): Atomic scale image of nanosheets. (b-e) Selected area electron diffraction patterns. (f) Atomic scale micrograph of phosphorene. (g) Interface between two sheets of phosphorene. Inset (iii): honey-comb microstructure of phosphorene.

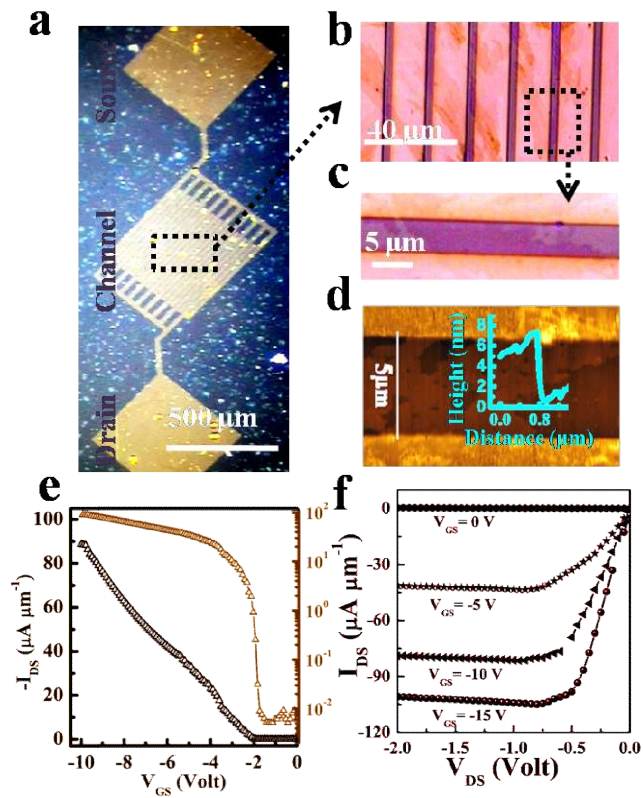

**Figure 4.** Field effect transistor with LB assembled phosphorene as a conducting channel material. (a) Digital image of the device. (b) Optical image of channel. (c) Magnified optical image showing a connecting nanosheet between source and drain. (d) AFM of the device. Inset: height profile of nanosheet. (e) Variation of drain current $I_{DS}$ with gate voltage $V_{GS}$ (left y-axis is the linear scale, and right y-axis is the logarithmic scale). (f) Variation of drain current $I_{DS}$ with the drain voltage $V_{DS}$ for different gate voltages $V_{GS}$.

*Supplementary Information*

*for*

# Large Area Fabrication of Semiconducting Phosphorene by Langmuir-Blodgett Assembly

Harneet Kaur[1], Sandeep Yadav[2], Avanish. K. Srivastava[1], Nidhi Singh[1], Jörg J. Schneider[2], Om. P. Sinha[3], Ved V. Agrawal[1,+], Ritu Srivastava[1,+,*]

[1]National Physical Laboratory, Council of Scientific and Industrial Research, Dr. K. S. Krishnan Road, New Delhi 110012, India.
[2]Technische Universität Darmstadt, Eduard-Zintl-Institut für Anorganische und Physikalische Chemie L2 I 05 117, Alarich-Weiss-Str 12, 64287 Darmstadt, Germany.
[3]Amity Institute of Nanotechnology, Amity University, Sector 125, Noida, Uttar Pradesh 201313, India.

*Corresponding Author (Email: ritu@nplindia.org)
[+]Authors contributed equally in this work.

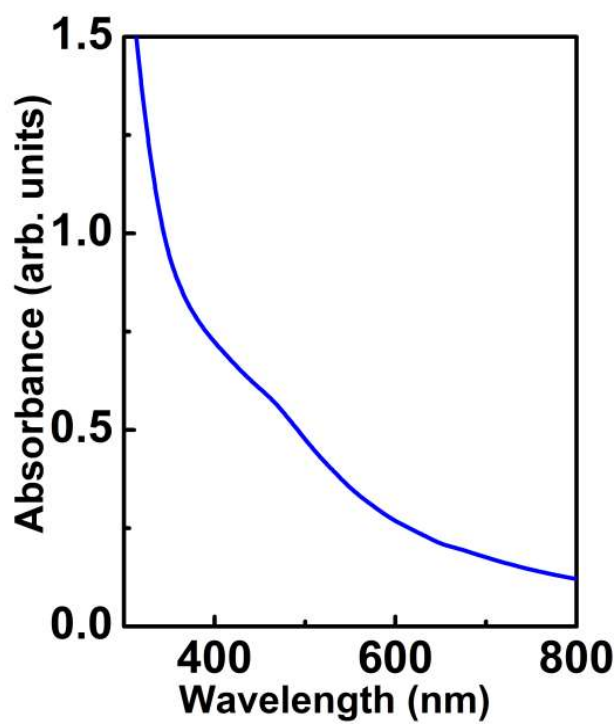

**Figure S1**: Optical absorbance spectra of exfoliated BP suspension in NMP, centrifuged at 3000 r.p.m.

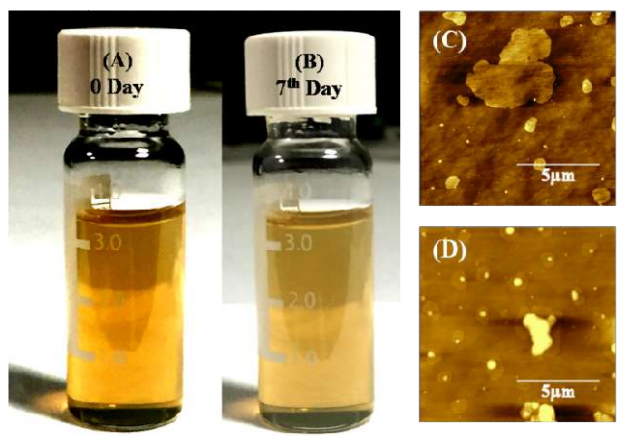

**Figure S2**: Digital image of phosphorene suspension and AFM of drop cast films. (A) Digital image on the first day of exfoliation. (B) Digital image after one week of ambient exposure showing the fading in the colour of suspension. (C) AFM of the drop-cast films prepared by using fresh exfoliated suspension. (D) AFM of the drop-cast films prepared by using one weak old suspension reveals the oxidation of nanosheets because of the formation of bubbles like structures on its surfaces.

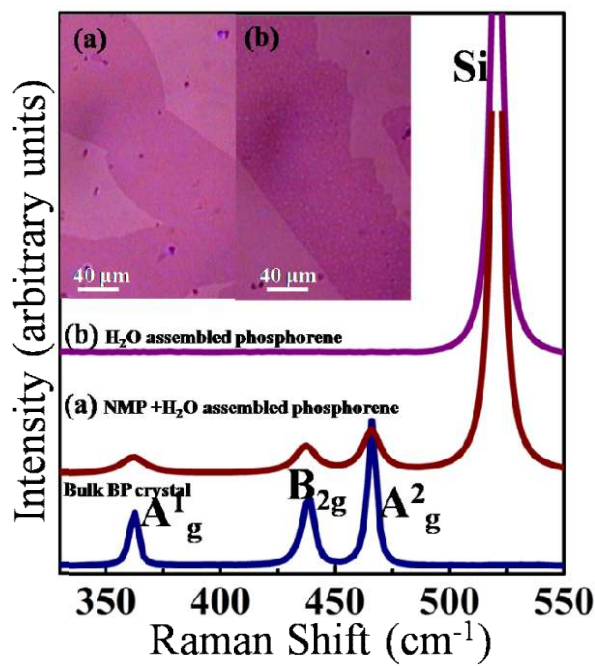

**Figure S3**: A comparison of the Raman spectra of bulk black phosphorus crystal (navy blue) and LB assembled phosphorene nanosheets (maroon and purple). The characteristic Raman modes are labelled. Inset (a): Confocal optical image of the nanosheet where Raman spectra was taken. These nanosheets were assembled using a mixture of NMP + deoxygenated water as subphase medium. The presence of Raman modes (maroon) and absence of bubbles in optical image suggests its pristine un-oxidized phase. Inset (b): Confocal

optical image of the nanosheet where Raman spectrum was taken. These nanosheets were assembled using deoxygenated water as a subphase medium. Presence of bubbles like structures on the surface of nanosheets in the optical image confirms its oxidation, resulting in absence of Raman vibrational modes (purple).

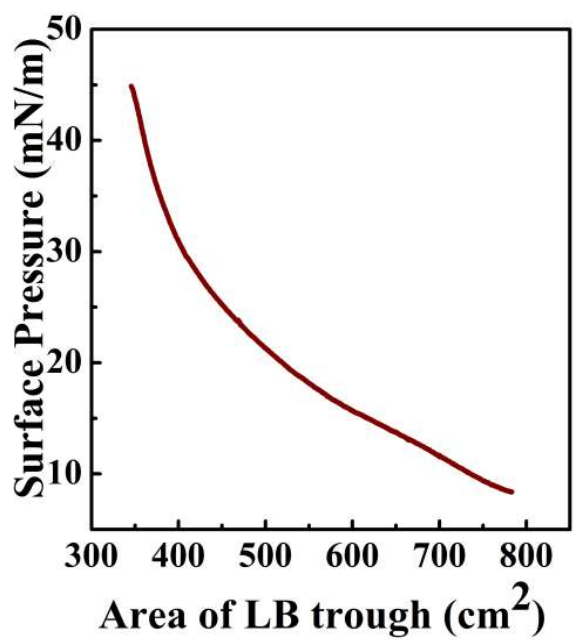

**Figure S4:** Variation in surface pressure with area of the LB trough on compressing the barriers. All the films were vertically lifted at a surface pressure of 40 mN/m.

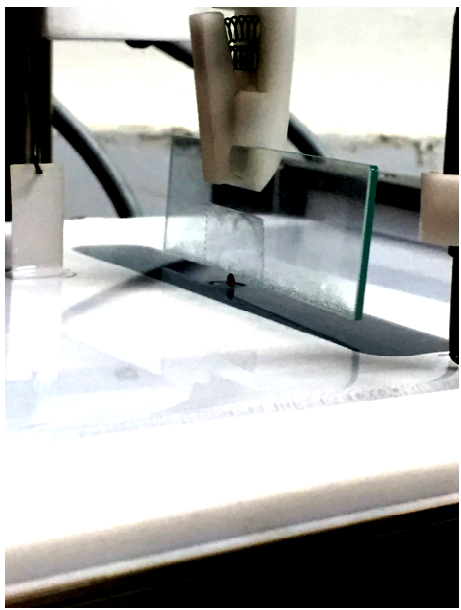

**Figure S5:** Digital image showing the vertical lift-off procedure on TEM grids in LB assembly.

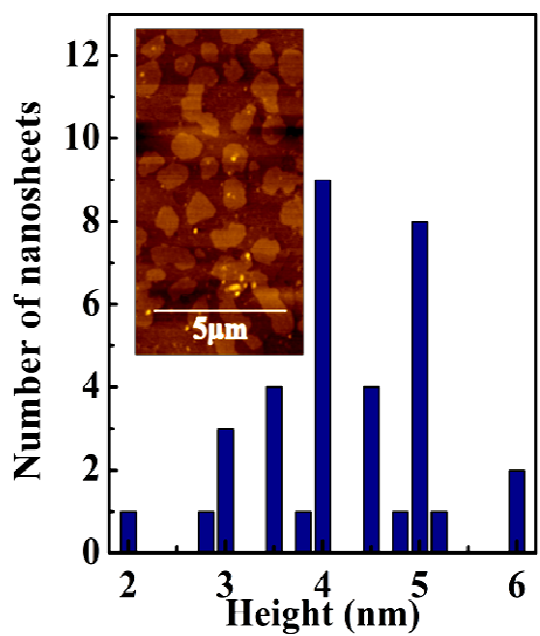

**Figure S6:** Height of the S-Ex BP nanosheets versus the number of sheets. Inset: AFM of the S-Ex BP nanosheets assembled by LB technique.

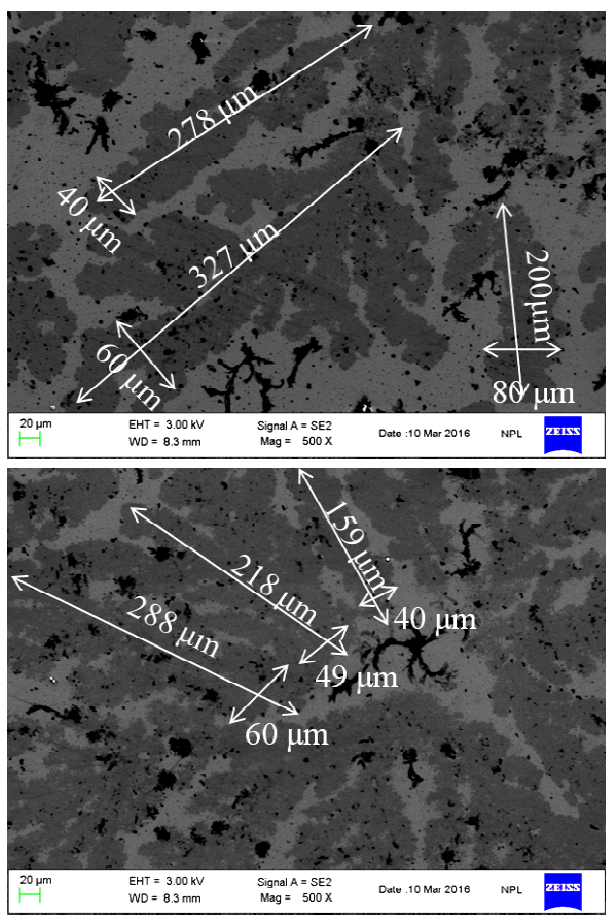

**Figure S7:** FESEM of LB assembled L-Ex BP nanosheets on SiO$_2$/Si substrate. The lateral dimension (defined as the largest side in our case) is of the order of hundreds of microns resulting in the enrichment of ultra-large nanosheets of phosphorene on substrate.